\documentclass[aps,twocolumn,prd,nofootinbib,showpacs]{revtex4}
\usepackage{amsmath}
\usepackage{amssymb}
\usepackage{graphicx}

\newcommand{\negfatspace}{\negthickspace \negthickspace}
\newcommand{\neghugespace}{\negfatspace\negfatspace}

\newcommand{\sss}[1]{{\scriptscriptstyle{#1}}}
\newcommand{\boldmathsymbol}[1]{\ensuremath{\boldsymbol{#1}}}

\newcommand{\vect}[1]{\boldmathsymbol{#1}}

\newcommand{\Gpc}{\textrm{Gpc}}
\newcommand{\Tb}{\textrm{Tb}}
\newcommand{\muK}{\mu\mathrm{K}}

\newcommand{\uCMB}{\mathrm{\sss{CMB}}}

\newcommand{\ures}{\mathrm{res}}

\newcommand{\uside}{\mathrm{side}}
\newcommand{\umat}{\mathrm{mat}}
\newcommand{\urad}{\mathrm{rad}}
\newcommand{\uhz}{\mathrm{hz}}
\newcommand{\ulz}{\mathrm{lz}}

\newcommand{\ud}{\mathrm{d}}
\newcommand{\ui}{\mathrm{i}}
\newcommand{\ue}{\mathrm{e}}

\newcommand{\OmegaM}{\Omega_\umat}
\newcommand{\OmegaR}{\Omega_\urad}

\newcommand{\U}{U}      
\newcommand{\GU}{G \U}
\newcommand{\Tcmb}{T_\uCMB}
\newcommand{\unitn}{\vect{\hat{n}}}

\newcommand{\boxgpc}{L_\mathrm{sim}}

\newcommand{\Xp}{\acute{X}}
\newcommand{\Xd}{\dot{X}}

\newcommand{\ang}{\theta}
\newcommand{\angres}{\ang_\ures}
\newcommand{\healpix}{\texttt{HEALPix}}
\newcommand{\mpi}{\texttt{MPI}}
\newcommand{\openmp}{\texttt{OpenMP}}
\newcommand{\hopper}{\texttt{hopper}}
\newcommand{\Nside}{N_\uside}
\newcommand{\Nsidehz}{N_\uside^{\uhz}}
\newcommand{\Nsidelz}{N_\uside^{\ulz}}

\begin{document}

\title{All sky CMB map from cosmic strings integrated Sachs-Wolfe
  effect}

\author{Christophe Ringeval} \email{christophe.ringeval@uclouvain.be}
\affiliation{Centre for Cosmology, Particle Physics and Phenomenology,
  \\ Institute of Mathematics and Physics, Louvain University, 2
  Chemin du Cyclotron, 1348 Louvain-la-Neuve, Belgium}

\author{Fran\c{c}ois R. Bouchet}
\email{bouchet@iap.fr}
\affiliation{Institut d'Astrophysique de Paris, UMR CNRS 7095,
  Universit\'e Pierre et Marie Curie, 98bis boulevard Arago, 75014
  Paris, France}

\date{\today}

\begin{abstract}
  By actively distorting the cosmic microwave background (CMB) over
  our past light cone, cosmic strings are unavoidable sources of
  non-Gaussianity. Developing optimal estimators able to disambiguate
  a string signal from the primordial type of non-Gaussianity requires
  calibration over synthetic full sky CMB maps, which untill now had
  been numerically unachievable at the resolution of modern
  experiments. In this paper, we provide the first high resolution
  full sky CMB map of the temperature anisotropies induced by a
  network of cosmic strings since the recombination. The map has about
  200 million sub-arcminute pixels in the healpix format which is the
  standard in use for CMB analyses ($\Nside=4096$). This premiere
  required about 800,000 cpu hours; it has been generated by using a
  massively parallel ray tracing method piercing through thousands
  of state of art Nambu-Goto cosmic string numerical simulations which
  pave the comoving volume between the observer and the last
  scattering surface. We explicitly show how this map corrects
  previous results derived in the flat sky approximation, while
  remaining completely compatible at the smallest scales.
\end{abstract}

\maketitle

\section{Introduction}
\label{sec:intro}

As all topological defects~\cite{Kibble:1976}, cosmic strings
incessantly generate gravitational perturbations all along the
universe history~\cite{Pen:1997, Durrer:1998ep, Zhou:1996}. Their
amplitude is directly given by the string energy density per unit
length, $G \U \ll 1$ (in Planck unit), which is also the typical
energy scale at which these objects are formed, eventually redshifted
by warped extra-dimension for the so-called cosmic
superstrings~\cite{Copeland:2003bj, Davis:2005, Sakellariadou:2009ev,
  Copeland:2009ga}. Although the theory of cosmological perturbations
can be applied to defects~\cite{Durrer:2002}, predicting string
induced anisotropies in the CMB is challenging. As opposed to the
perturbations of inflationary origin, which are generated once and for
all in the early universe, active sources require the complete
knowledge of their evolution at all times, from their formation till
today~\cite{Pen:1994, Magueijo:1996px, Magueijo:1996xj,
  Achucarro:1998ux, Contaldi:1998mx}.

For these reasons, cosmological analyses often rely on analytical, or
semi-analytical defect models~\cite{Pogosian:1999np, Gangui:2001fr,
  Siemens:2006vk, Jeong:2007, Pogosian:2007gi, Pshirkov:2009vb,
  Danos:2009vv, Tuntsov:2010fu, Danos:2010gx, Demorest:2012bv,
  Kamada:2012ag} which may not be accurate enough in view of the
incoming flow of high precision CMB data, such as those from the
Planck satellite and the other sub-orbital
experiments~\cite{Huffenberger:2005, Ruhl:2004, Ami:2006,
  2011AA536A1P}. The theoretical understanding of cosmic string
evolution in a Friedmann-Lema\^{i}tre-Robertson-Walker (FLRW) universe
is still an active field of research which has led to the development
of various theoretical models~\cite{Austin:1993rg, Copeland:1998na,
  Martins:2000cs, Polchinski:2006ee, Dubath:2007mf, Rocha:2007ni,
  Copeland:2009dk, Martins:2009hj, Lorenz:2010sm, Pourtsidou:2010gu,
  Avgoustidis:2011ax} and numerical simulations, the latter having the
advantage of incorporating all the defect dynamics~\cite{Bennett:1989,
  Albrecht:1989, Bennett:1990, Allen:1990, Vincent:1998, Moore:2002,
  Wu:2002, Ringeval:2005kr, Vanchurin:2005yb, Olum:2006ix,
  Martins:2005es, BlancoPillado:2011dq}. However, within simulations,
the dynamical range remains limited such that one has to extrapolate
numerical results over orders of magnitude by means of their scale
invariant properties. The CMB temperature and polarization angular
power spectra have been derived for local strings only within the
Abelian Higgs model~\cite{Bevis:2006mj, Bevis:2007qz,
  Bevis:2010gj}. As simulations do have only one free parameter that
is $G \U$, they provide a robust correspondence between string tension
and CMB amplitude. From current data, Ref.~\cite{Urrestilla:2011gr}
reports the two-sigma confidence limit $\GU< 4.2 \times 10^{-7}$,
whereas semi-analytical methods find bounds ranging from $10^{-9}$ to
$10^{-6}$~\cite{Fraisse:2006xc, Movahed:2010zq, Battye:2010xz,
  Dvorkin:2011aj}.

Among other signatures, non-Gaussianities are unavoidable consequences
of the presence of cosmic strings (see Ref.~\cite{Ringeval:2010ca,
  Hindmarsh:2011qj} for a review). The determination of the power
spectrum, i.e. the two-points function, from numerical simulations
is not easy, and the situation is even worse for any higher n-point
function. A way around this is to include photons inside a string
simulation to produce a realization of the expected CMB temperature
anisotropies. In that respect, the resulting map contains all the
statistical content, non-Gaussianities included. This method has
originally been introduced for Nambu-Goto strings in
Ref.~\cite{Bouchet:1988} and revived in Ref.~\cite{Fraisse:2007nu} to
create a collection of statistically independent small angle CMB
maps. As shown by Hindmarsh, Stebbins and Veeraraghavan, the small
angle limit happens to be very convenient as the perturbed photon
propagation equations, namely the string induced integrated
Sachs-Wolfe (ISW) effect~\cite{Gott:1984ef, Kaiser:1984}, reduces to a
more tractable two-dimensional problem~\cite{Hindmarsh:1993pu,
  Stebbins:1995}. Those maps have been shown to be accurate as they
correctly reproduce the small scale power spectrum of Abelian
strings~\cite{Bevis:2010gj} as well as the analytically expected
one-point~\cite{Takahashi:2008ui, Yamauchi:2010ms, Yamauchi:2010vy}
and higher n-point functions~\cite{Hindmarsh:2009qk,
  Hindmarsh:2009es, Regan:2009hv}. Although flat sky maps are adequate
to devise new string-oriented searches in small scale CMB
data~\cite{Hammond:2009, Niemack:2010wz, Dunkley:2010ge}, current
searches for non-Gaussianities are mainly driven by the primordial
type and based on full sky optimized estimators~\cite{Liguori:2010hx,
  Regan:2010cn, Fergusson:2009nv, Curto:2011zt} (see
Ref.~\cite{Huterer:2010en} for a review).

In this paper, we generalize the method used in
Ref.~\cite{Fraisse:2007nu} and go beyond the flat sky approximation to
generate a full sky CMB map induced by Nambu-Goto strings. The most
unequivocal signature from strings are the temperature discontinuities
they induce, which are naturally most striking at small scales; our
efforts thus have been to achive a high resolution over the complete
sky. In a hierarchical equal area isolatitude pixelisation of the sky,
we have been able to maintain an angular resolution of
$\angres=0.85'$, i.e. using the publicly available $\healpix$
code~\cite{Gorski:2004by}, our map has $\Nside=4096$, i.e.  $2\times
10^8$ pixels. As detailed in the following, our method includes all
string effects from the last scattering surface till today, but does
not include the Doppler contributions induced by the strings into the
plasma prior to recombination. As a result, our map represents the ISW
contribution from strings, which is dominant at small scales but
underestimates the signal on intermediate length scales. Including the
Doppler effects requires the addition of matter in the simulations, an
approach which has been implemented in Refs.~\cite{Landriau:2002fx,
  Landriau:2003xf} and recently used to generate a full sky map in
Ref.~\cite{Landriau:2010cb}. As discussed in that reference, the
computing resources required to include matter severely limit the
achievable resolution to $14'$ ($\Nside=256$, with 800 000 pixels). In
that respect, our map is complementary to the one of
Ref.~\cite{Landriau:2010cb} while extending the domain of
applicability of existing small angle maps~\cite{Fraisse:2007nu}. In
particular, we recover the turnover in the spectrum observed
around $\ell\simeq 200$ in Ref.~\cite{Landriau:2010cb}, which was cut
by the small field of view of the flat sky maps.

The paper is organized as follows. In Sec.~\ref{sec:simu}, we briefly
recap the characteristics of our numerical simulations of cosmic
string networks and describe the ray tracing method used to compute
the full sky map. The map itself is presented in Sec.~\ref{sec:map}
and compared with the small angle maps in the applicable limit. We
also compute the angular power spectrum and conclude in the last
section.

\section{Method}
\label{sec:simu}

\subsection{All sky string ISW}

Denoting by $X^\mu(\tau,\sigma)$ the string embedding functions, in
the transverse temporal gauge\footnote{$\vect{\Xd}\cdot\vect{\Xp}=0$,
  $X^0=\eta=\tau$, $\eta$ is the conformal time.}, up to a dipole
term, the integrated Sachs--Wolfe contribution sourced by the
Nambu-Goto stress tensor reads~\cite{Stebbins:1995}
\begin{equation}
\label{eq:isw}
\Theta(\unitn) = - 4 G \U \int_{\vect{X}\,\cap\,\vect{x}_\gamma} \neghugespace
\vect{u} \cdot \dfrac{X \unitn - \vect{X}}{\left(X \unitn -
    \vect{X} \right)^2} \,\epsilon\, \ud \sigma,
\end{equation}
where $\Theta(\unitn)$ stands for the relative photon temperature
shifts ($\Theta \equiv \Delta T/T$) in the observation direction
$\unitn$. The integral is performed over all string position vectors
$\vect{X}=\{X^i\}$ intercepting our past line cone, the invariant
string length element being $\ud l=\epsilon \ud \sigma$ with
$\epsilon^2 \equiv \vect{\Xp}^2/(1-\vect{\Xd}^2)$. The string
dynamical effects are encoded in
\begin{equation}
\label{eq:udef}
\vect{u} = \vect{\Xd} - 
  \dfrac{(\unitn \cdot \vect{\Xp}) \cdot \vect{\Xp}}{1 +
  \unitn \cdot \vect{\Xd}},
\end{equation}
where the ``acute accent'' and the ``dot'' denote differentiation with
respect to the world sheet coordinates $\sigma$ and $\tau$,
respectively. The small angle and flat sky approximations consists in
taking the limit $X\unitn \rightarrow \vect{X}$ which assumes that the
observation direction matches with the string location and that all
photon trajectories are parallel~\cite{Hindmarsh:1993pu, Bouchet:1988,
  Fraisse:2007nu}. Let us notice that Eq.~(\ref{eq:isw}) cannot be
applied to a straight static string, but such a situation never occurs
for the realistic string configurations studied in the
following~\cite{Stebbins:1995, 1992ApJ395L55V}.

In the following, we use the all sky expression of Eq.~(\ref{eq:isw})
to compute the induced temperature anisotropies $\Theta(\unitn)$ in
each of the wanted $2\times 10^{8}$ pixelized
directions. Equation~(\ref{eq:isw}) shows that, in each direction, one
has to sum up the contribution of \emph{all} string segments $\ud l$
intercepting our past light cone since the last scattering surface and
determine, for each of them, $\vect{\Xp}$ and $\vect{\Xd}$. Although
one string lying behind the observer does not contribute more than a
few percent to the overall signal, it is impossible to artificially
cut it without adding spurious discontinuities in the map, which would
dangerously mimic real string patterns. As discussed below, we have
filled the comoving volume between today and the last scattering
surface with a few thousand Nambu-Goto numerical simulations in
Friedmann-Lema{\^i}tre-Robertson-Walker (FLRW) space-time. Typically,
hundreds of millions of string segments intercept on our past light
cone, each of them has to be included in Eq.~(\ref{eq:isw}) to get the
overall signal for one pixel. One can immediately understand the
computing challenge to obtain a full sky map as the total number of
expected iterations roughly sums up to $10^{16}$.

\subsection{Nambu-Goto string simulations}

\label{sec:ngsim}

In order to get a realistic string configuration between the last
scattering surface and today, we have followed
Refs.~\cite{Bouchet:1988, Fraisse:2007nu} and stacked FLRW string
simulations using an improved version of the Bennett and Bouchet
Nambu-Goto cosmic string code~\cite{Bennett:1990,
  Ringeval:2005kr}. The runs have the same characteristics as those
used in Ref.~\cite{Fraisse:2007nu} and, in particular, we include only
the loops having a size larger than a time-dependent cut-off. The
reason being that loops smaller than this cut-off have a distribution
which is known to be contaminated by relaxation effects from the
numerical initial conditions. As explained in
Ref.~\cite{Fraisse:2007nu}, the cutoff is dynamically chosen by
monitoring the time evolution of the energy density distribution
associated with loops of different sizes. This ensures that we include
only loops having an energy density evolving as in scaling, i.e. in
$1/t^2$. One may be worried about the deficit due to the missing loops
artificially removed by the cut-off. An upper bound of the systematic
errors that could be induced by these relaxation effects can be found
in that same reference (see Sec.~II.D). It does not exceed $10\%$ on
average and concerns only very small scales. In the following, we
recap some of the relevant physical properties underlying our
numerical simulations (see Sec.~II.B in Ref.~\cite{Fraisse:2007nu} for
more details). Each simulation allows one to trace the time evolution of a
network of cosmic strings in scaling over a cubic comoving volume of
typical size
\begin{equation}
\label{eq:boxgpc}
\boxgpc \simeq 30 \dfrac{1 -
  \sqrt{\OmegaR/\OmegaM} \sqrt{1+z_\ui}}
{h \sqrt{\OmegaM} \sqrt{1 + z_\ui}} \,\Gpc.
\end{equation}
In this expression, $z_\ui$ is the redshift at which the simulation is
started, $h$ is the reduced Hubble parameter today and $\OmegaM$,
$\OmegaR$ are the density parameters of matter and radiation
today. Starting at the last scattering surface, $z_\ui=1089$, gives a
simulation comoving box of $\boxgpc \simeq 1.7\,\Gpc$\footnote{In the
  runs, $\boxgpc$ is computed exactly within the $\Lambda$CDM model
  whereas Eq.~(\ref{eq:boxgpc}) is an analytical approximation
  assuming no cosmological constant and $\OmegaR \ll \OmegaM$.}, which
corresponds to an angular size of $7.2^\circ$. At the same time
strings are evolved, we propagate photons along the three spatial
directions and record all $\vect{\Xp}$ and $\vect{\Xd}$ for each
string segment intercepting those light cones. Depending on the
simulation realization, and its location, this corresponds to
typically $10^4$--$10^5$ projected string segments. One run is limited
in time, as we use periodic boundary conditions, and ends after a
30-fold increase in the expansion factor, i.e. at a redshift
$z_\ue=36$. As a result, covering the whole sky requires stacking
side-by-side many different simulations, all starting at $z_\ui=1089$
and ending at $z_\ue=36$. The missing redshift range can be dealt
exactly in the same manner by stacking another set of runs which start
at $z_\ui = 36$ and end at $z_\ue= 0.2$. From Eq.~(\ref{eq:boxgpc}),
we see that the low-redshift simulations have a size of $\boxgpc
\simeq 13 \,\Gpc$ such that only a few of them are required to cover
the whole comoving volume. Notice that the use of different
simulations to fill the comoving space does not induce visible
artifacts in the final map. The signal being only sourced by the
subset of string segments intercepting the past light cone, the
probability of seeing an edge is almost vanishing. Finally, as in
Ref.~\cite{Fraisse:2007nu}, we have skipped the last interval from
$z=0.2$ to $z=0$ as almost no string intercepts our past line cone in
that range.

In the next section, we describe in more detail how we cut and stack
the cosmic string simulations.

\subsection{Stacking simulations with healpix cones}

\begin{figure}
\begin{center}
\includegraphics[width=\columnwidth]{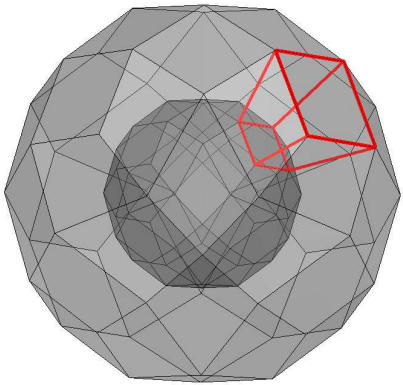}
\caption{Voxelization of the comoving space between the two redshifts
  at which cosmic string simulations start and stop, here $z_\ui=1089$
  (outer sphere) and $z_\ue=36$ (inner sphere). All strings living in
  each healpix cone are kept and stacked to fill the whole comoving
  volume. Our high redshift voxelization scheme uses $\Nsidehz=16$,
  i.e. requires $3072$ string simulations while the low redshift one,
  starting at $z_\ui=36$ and ending now, has $\Nsidelz=1$, i.e. $12$
  simulations.}
\label{fig:healvox}
\end{center}
\end{figure}

The hierarchical equal area iso-latitude pixelisation (healpix) of the
two-sphere is a efficient method to pixelize the sky which is
well-suited to and commonly used for CMB
analysis~\cite{Gorski:2004by}. Our stacking method relies on the
voxelization of the three-dimensional ball using cones subtending
healpix pixels on its boundaries, i.e. for the two redshifts at which
we start and stop the numerical simulations. For the high-redshift
contributions, the healpix cones fill the whole spherical volume in
between the last scattering surface at $z_\ui=1089$ and the two-sphere
at $z_\ue=36$ (see Fig.~\ref{fig:healvox}). The rest of the volume,
associated with the low redshift simulations, can be voxelized exactly
in the same way but starting on the two-sphere at $z_\ui=36$ and
ending at $z_\ue=0.2$. The actual string dynamics simulations are
evolved in cubic comoving boxes in which we take only strings living
inside a healpix cone. In order to minimize the number of simulations
required, the problem is now reduced to find the largest healpix cone
fitting inside a cubic comoving box for all redshifts of
interest. Moreover, one has to ensure that the photons intercepting
the strings travel towards the observer. As seen in
Fig.~\ref{fig:healvox}, both requirements can be implemented by
adequately rotating the simulation box such that photons face the
observer line of sight $\unitn$, and the farthest healpix pixel fits
inside the farthest squared face of the simulation box. As we
propagate three photon waves in each simulation, one can use the same
simulation rotated three times. Keeping only the strings living inside
the healpix cones, one needs to patch them up till the whole comoving
volume is filled. For this purpose, we have used the algorithms
implemented within the $\healpix$ library~\cite{Gorski:2004by}.

For the high-redshift contribution, the above requirements are
satisfied with a healpix voxelization having $\Nsidehz=16$,
i.e. $3072$ truncated cones from $z=1089$ to $z=36$, therefore calling
for $3072$ cosmic string runs. As we can use one run three times, we
have only performed $1024$ independent simulations as described in
Sec.~\ref{sec:ngsim} (and Ref.~\cite{Fraisse:2007nu}). Concerning the
low-redshift contribution, the simulation volume $\boxgpc$ being much
larger, $\Nsidelz=1$ or $12$ simulations are enough to cover the
volume. Let us emphasize, at this stage, that the above-mentioned
$\healpix$ resolutions only concern the simulation stacking method. In
the next section, we discuss how we discretize the CMB sky using
another, but more conventional, healpix scheme.

\subsection{Healpix sky}

Assuming we have stacked the cosmic string simulations as previously
explained, we have at our disposal a realization of all $\vect{\Xd}$
and $\vect{\Xp}$ lying on our past light cone since the last
scattering surface. From Eqs.~(\ref{eq:isw}) and (\ref{eq:udef}), the
remaining step is to actually perform the integral for each desired
value of the observer direction $\unitn$. Choosing the values of
$\unitn$ has been made by using another healpix pixelization scheme,
this time on the simulated sky. A typical Planck-like CMB experiment
requiring an angular resolution of $5'$, we have chosen an angular
resolution of $0.9'$ to reduce sufficiently the small scale
inaccuracies which are present at the subpixel scale. The
corresponding healpix resolution is $\Nside=4096$, i.e. a sky map
having $2 \times 10^8$ pixels.

In the next section, after having briefly exposed how the above
computing challenges have been solved, we present the simulated CMB
map.

\section{String sky map}
\label{sec:map}

The method we have exposed in the previous section has the advantage
to be completely factorizable into two independent numerical
problems. The first is to perform string simulations and record only
those events intercepting the light cones. The second is to use these
events to actually perform the integration of Eq.~(\ref{eq:isw}),
which gives the final signal $\Theta(\unitn)$.

\subsection{Computing resources}

Performing the one thousand cosmic string runs has required around
$300,000$ cpu-hours on current x86-64 processors. The needed memory and
disk space resources remaining reasonable, the computations have used
local computing resources provided by the Planck-HFI computing
centre at the ``Institut d'Astrophysique de Paris'' and the CP3 cosmo
cluster at Louvain University. By the end of the runs, the total
number of string segments recorded on the light cone account for
typically $1\, \Tb$ of storage data.

From the light cone data, performing $2 \times 10^8$ line of sight
integrals over $10^8$ string segments [see Eq.~(\ref{eq:isw})] is a
serious numerical problem. This part of the code has therefore been
parallelized at three levels using the symmetries of the problem.

First, we have used a distributed memory parallelization as
implemented in the message passing interface ($\mpi$) to split the
string contributions into sub-blocks. From our previous discussing, a
natural implementation is an MPI-parallelization over the healpix cones
which therefore allows different machines to deal with a subset of
cones. More intuitively, it means that the final map is obtained by
adding up full sky maps, each one being sourced by the strings
contained in a few healpix cones only. Second, for each of those
processes, the computation of the $2\times 10^8$ pixels has been
parallelized using the shared memory $\openmp$ directives. In other
words, pixels can be simultaneously computed using all of the
available processors inside a single machine. Finally, for each of the
above $\openmp$ threads, we have vectorized the most inner loop,
i.e. the discrete version of Eq.~(\ref{eq:isw}). Doing this allows
to use simultaneously multiple registers of each processing unit to
add up a few string segments at once.

The computing resources have been provided by the National Energy
Research Scientific Computing Center (NERSC) at the Lawrence Berkeley
National Laboratory\footnote{\texttt{http://www.nersc.gov}}. The run
has been performed on the $\hopper$ super-computer using $512$ $\mpi$
nodes at the first level of parallelization. On each node, we have
deployed the $\openmp$ parallelization over the $24$ cores
available. Each node is made of two twelve-cores ``AMD MagnyCours''
cpus which supports only a limited amount of vectorization. However, a
vectorization over $16$ string segments significantly improves cache
memory latency and gives a final speed-up of two compared to a pure
scalar processing. All in all, the whole computation required $12,000$
cores and has been completed after $500,000$ cpu-hours.

\subsection{Results}

\begin{figure*}
\begin{center}
\includegraphics[angle=90,width=0.65\textwidth]{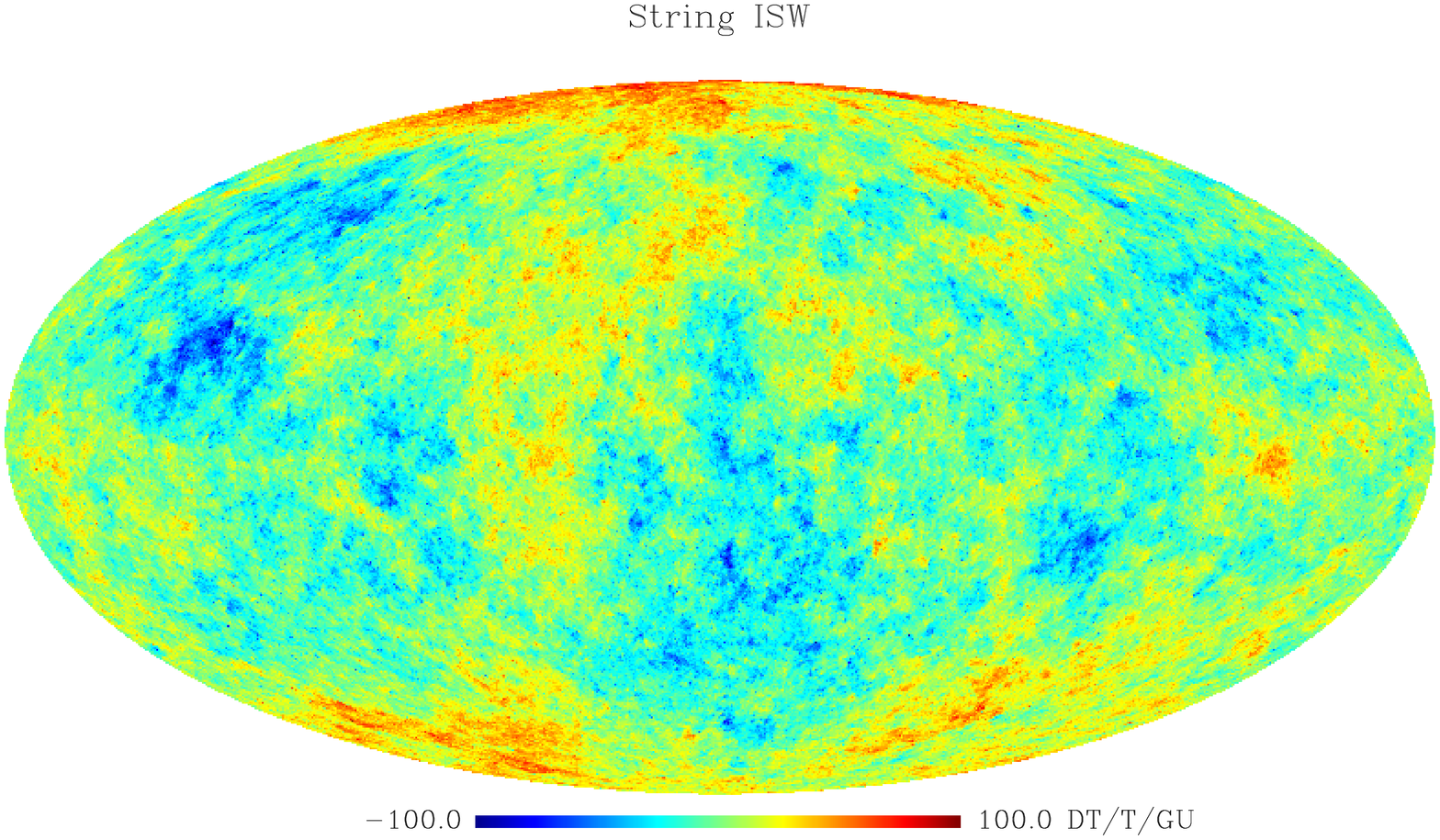}
\includegraphics[angle=90,width=0.65\textwidth]{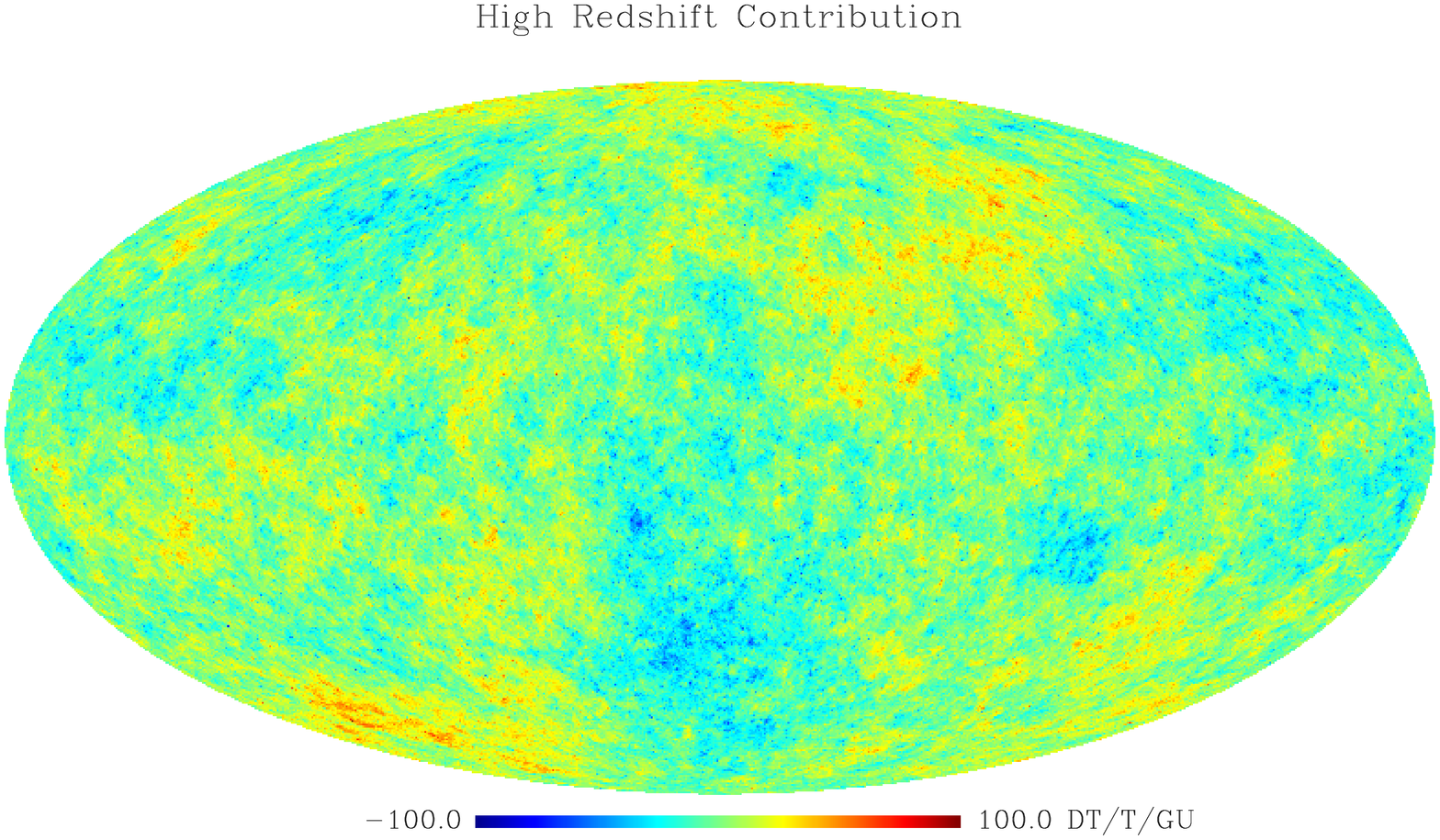}
\includegraphics[angle=90,width=0.65\textwidth]{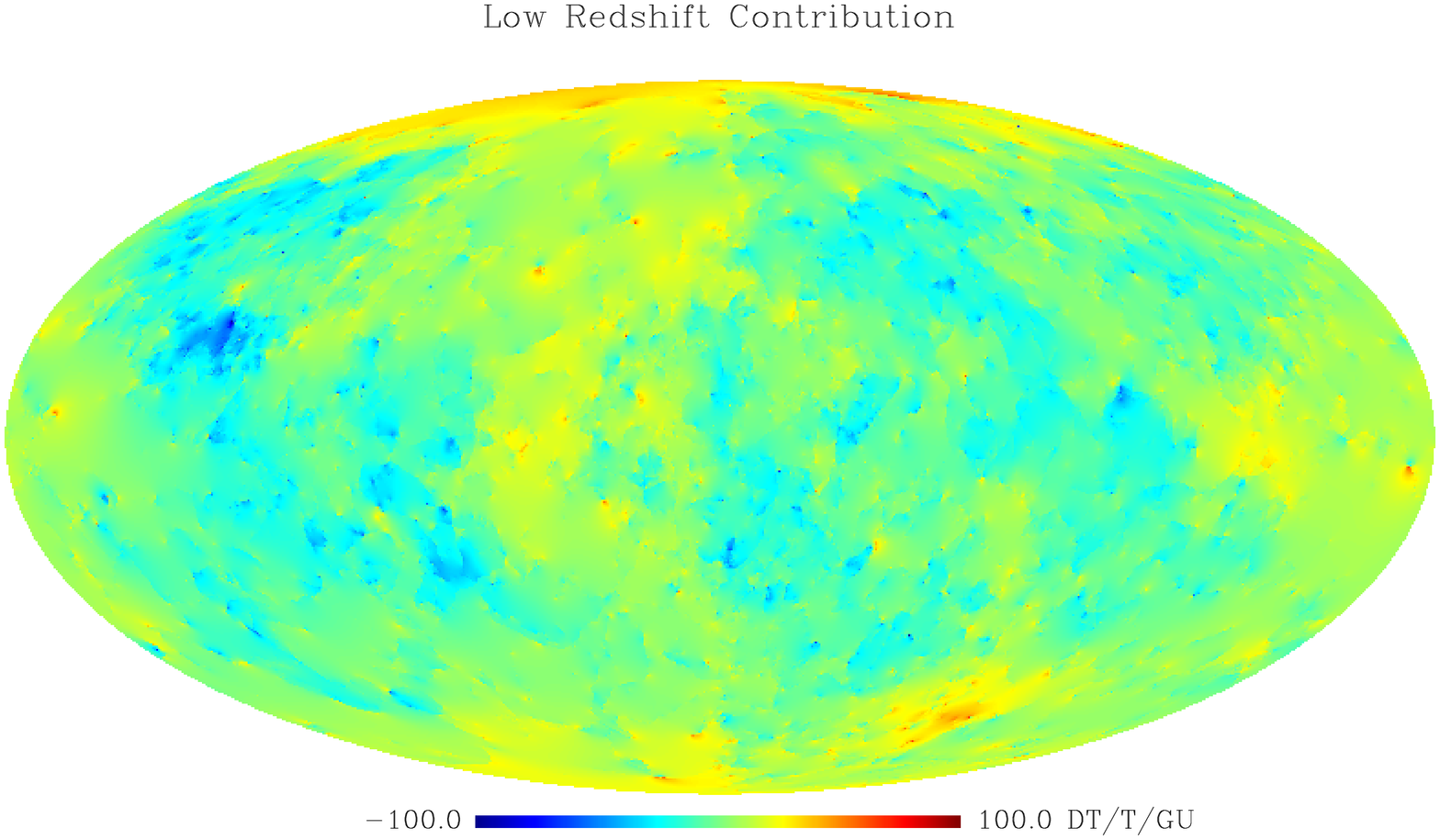}
\caption{All sky CMB map of string induced ISW effect (upper panel) in
  a Mollweide projection. The two lower maps represent respectively
  the high redshift contribution (from last scattering to $z=36$) and
  the low redshift contribution (from $z=36$ to now). The color scale
  indicates the range of $\Theta/GU$ fluctuations which, once
  multiplied by the value of $\GU$ and $\Tcmb$, would thus be in the tens of
  $\muK$ range for $\GU \sim 10^{-6}$.}
\label{fig:maps}
\end{center}
\end{figure*}

In Fig.~\ref{fig:maps}, we have plotted the final map together with
the two contributions coming from high and low redshift. As expected,
most of the strings are at high redshift, the low redshift
contribution showing only a few strings crossing the sky. For this
reason, we have not included strings with $z<0.2$, as almost no strings
are present.

\begin{figure}
\begin{center}
\includegraphics[width=0.88\columnwidth]{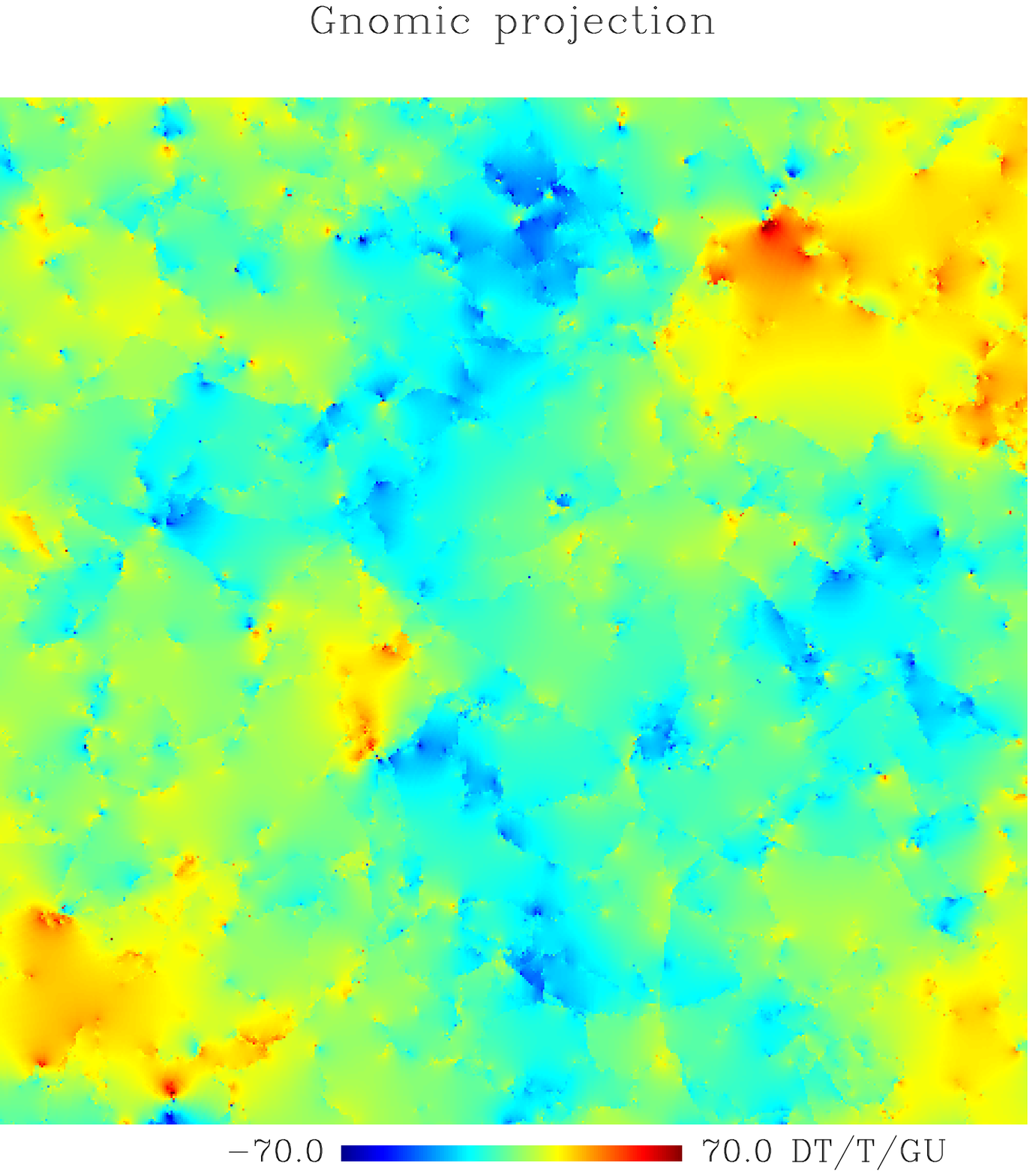}
\includegraphics[width=0.88\columnwidth]{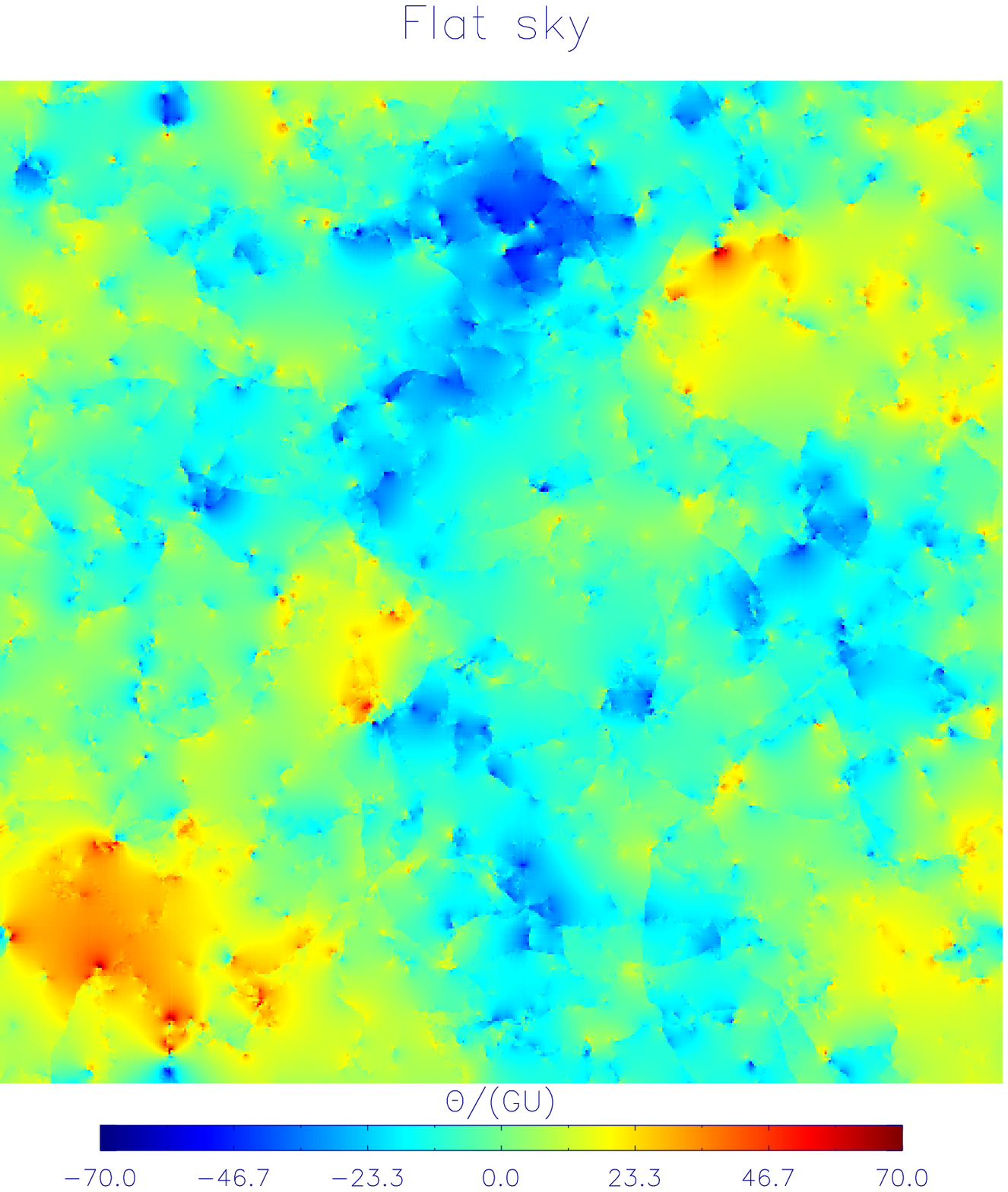}
\caption{Comparison between full sky and flat sky / small angle
  approximation. The upper panel is a gnomic projection of a
  $7.2^\circ$ patch cut from the high redshift full sky map at the
  north pole. The lower panel is from Ref.~\cite{Fraisse:2007nu} and
  represents the high redshift flat sky calculation coming from the
  same string simulation. Both maps exhibit globally the same
  anisotropy patterns and amplitude. More precisely, there is
  essentially no difference in the center, whereas geometrical and
  amplitude distortions become increasingly apparent towards the
  edges. This is expected as spherical effects are not included in the
  flat sky approximation.}
\label{fig:cuts}
\end{center}
\end{figure}

At first glance these maps may look Gaussian, which is because the
string patterns essentially show up at small scales while they are
averaged on the largest angular scales~\cite{Landriau:2010cb}. In
Fig.~\ref{fig:cuts}, we have represented a zoom over a $7.2^\circ$
region in which one recovers the same string discontinuities as
previously derived within the flat sky
approximation~\cite{Fraisse:2007nu}. In order to make the comparison
sharper, both the gnomic projection of our spherical patch and the
flat sky map coming from the same string simulation are represented in
Fig.~\ref{fig:cuts}. Up to some spherical distortions off-center, both
maps predict the same structures.

\begin{figure}
\begin{center}
\includegraphics[width=0.9\columnwidth]{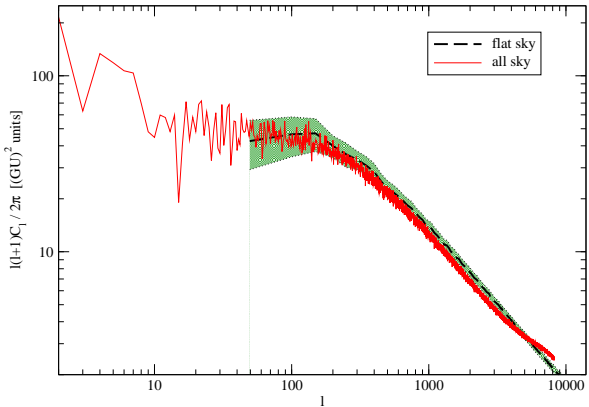}
\caption{Angular power spectrum extracted from the string map of
  Fig.~\ref{fig:maps} (solid red curve) compared to the one estimated
  in Ref.~\cite{Fraisse:2007nu}, using the small angle and flat sky
  approximation (dashed black curve). The shaded region represents the
  one-sigma fluctuations over various string realizations (flat sky).  }
\label{fig:ps}
\end{center}
\end{figure}

Finally, in Fig.~\ref{fig:ps}, we have plotted the angular power
spectrum obtained from the full sky map. In this figure, the dashed
curve represents the mean value previously derived using flat sky
patches in Ref.~\cite{Fraisse:2007nu}. We observe a very small loss of
power compared to the flat sky result, which may not be significant as
it does not exceed the one-sigma fluctuations expected between various
string realizations. However, as Fig.~\ref{fig:cuts} suggests, the
flat sky maps are a bit sharper than the spherical ones due to some
missing spherical effects and, as such, they may slightly overestimate
the signal. This is not surprising as the flat sky maps are derived
using the small angle approximated version of Eq.~(\ref{eq:isw}) where
all scalar products are postulated to be either parallel or
null~\cite{Hindmarsh:1993pu}. As our full sky map does not approximate
anything, it should contain slightly less power.

Let us also notice that there is some extra power for $\ell>5000$
associated with the full sky power spectrum. This is a spurious aliasing
effect coming from the slow decrease of the power spectrum at small
scales combined with our ray tracing method. Each pixel of the map
represents the real signal, but only in the centroid direction
$\unitn$. This has the effect of including string structures smaller
than the pixel angular resolution thereby aliasing the final map.

In Fig.~\ref{fig:ps}, the power spectrum turns-over for $\ell<200$ as
those modes correspond to wavelengths larger than the correlation
length at recombination. This property was also observed in the map
derived in Ref.~\cite{Landriau:2010cb} but barely visible in the flat
sky maps of Ref.~\cite{Fraisse:2007nu} due to their small field of
view. We see, a posteriori, that this effect was however indeed
present as the dash curve of Fig.~\ref{fig:ps} flattens at the same
location than the full sky spectrum.

\section{Conclusion}

The main result of this article is the map displayed in
Fig.~\ref{fig:maps} which provides a realization of the all sky CMB
temperature anisotropies induced by a network of cosmic strings since
the last scattering surface. The challenges underlying this map come
from both the requirement of covering the whole sky and an
unprecedented angular resolution of $0.85$ minute of arc, associated
with a $\healpix$ resolution of $\Nside=4096$. However our map
includes only the ISW string contribution. This is the dominant signal
at small scales, but it misses the Doppler effects around the
intermediate multipoles. As those effects have been computed in
Ref.~\cite{Landriau:2010cb}, at the expense of having a poor
resolution, it will be interesting to investigate whether both results
can be combined to obtain a fully accurate representation of the
stringy sky over the full range of observable scales.

\acknowledgments

This research used resources of the National Energy Research
Scientific Computing Center, which is supported by the Office of
Science of the U.S. Department of Energy under Contract
No. DE-AC02-05CH11231. This work is also partially supported by the
Wallonia-Brussels Federation Grant $\mathrm{N}^\circ$ ARC 11/15-040
and ESA under the Belgian Federal PRODEX Program $\mathrm{N}^\circ
4000103071$.

\bibliography{strings}

\end{document}